\documentclass[journal]{IEEEtran}
\usepackage{graphicx} 
\usepackage{comment}
\usepackage{booktabs}
\usepackage{makecell}
\usepackage{caption}
\usepackage{xcolor}
\usepackage{subfigure}
\usepackage{multirow}
\usepackage{textcomp}
\usepackage{cite}

\usepackage{graphicx}

\usepackage{amsmath}

\usepackage{algorithmic}

\usepackage{array}

\ifCLASSOPTIONcompsoc
  \usepackage[caption=false,font=normalsize,labelfont=sf,textfont=sf]{subfig}
\else
  \usepackage[caption=false,font=footnotesize]{subfig}
\fi

\usepackage{url}

\begin{document}

\title{Covert Association of Applications on Edge Devices by Processor Workload}

\author{
  \IEEEauthorblockN{Hangtai~Li $\qquad$ Yingbo~Liu $\qquad$ Rui~Tan}\\
  \IEEEauthorblockA{School of Computer Science and Engineering, Nanyang Technological University, Singapore\\
  \{N1801735A, liuyb, tanrui\}@ntu.edu.sg}
  }

\maketitle

\begin{abstract}
  The scheme of application (app) distribution systems involving incentivized third-party app vendors is a desirable option for the emerging edge computing systems. However, such a scheme also brings various security challenges as faced by the current mobile app distribution systems. In this paper, we study a threat named {\em covert device association}, in which the vendors of two apps collude to figure out which of their app installations run on the same edge device. If the two colluding apps are popular, the threat can be used to launch various types of further attacks at scale. For example, the user of the compromised edge device, who wishes to remain anonymous to one of the two apps, will be de-anonymized if the user is not anonymous to the other app. Moreover, the coalition of the two apps will have an escalated privilege set that is the union of their individual privilege sets. In this paper, we implement the threat by a reliable and ubiquitous covert channel based on the edge device's processor workload. The implementations on three edge devices (two smartphones and an embedded compute board) running Android and Android Things do not require any privileged permissions. Our implementations cover three attack scenarios of 1) two apps running on the same Android phone, 2) an app and a web session in the Tor browser running on the same Android phone, and 3) two apps running on the same Android Things device. Experiments show that the covert channel gives at least 0.25 bps data rate and the covert device association takes at most 3.2 minutes.
\end{abstract}

\IEEEpeerreviewmaketitle

\section{Introduction}
\label{sec:intro}

The Internet has been fast expanding its edge to interconnect a wide spectrum of smart devices (e.g., embedded sensors, smartphones, wearables, and home appliances) and form the Internet of Things (IoT). It is estimated that, by 2019, there are about 26.66 billion IoT devices \cite{IoT2019darina}. This quantity is predicted to be doubled by 2022 \cite{statista2019}. The IoT devices can be roughly categorized into two classes of {\em end devices} and {\em edge devices}. An end device is often a smart version of a traditional special-purpose physical object. As the end devices do not provide generic computing services, their computing units are often customized by the manufacturers. Various smart home appliances are representative examples, such as smart lights, toothbrushes, body scales, etc. Differently, edge devices often provide some form of generic computing services and can serve as the intermediaries between the end devices and the cloud backend. Examples include smartphones (as mobile edges), set-top boxes, and home gateways.

The current prevalent mobile operating systems (OSes), e.g., Android and iOS, have fostered thriving application (app) distribution systems involving large communities of third-party app vendors and users. Such an app-centric model is being adopted by emerging OSes that target a broader scope of edge devices, such as Ubuntu IoT, Windows IoT, and Android Things. Different from end devices that mostly run device manufacturers' firmwares only, edge devices face a unique challenge of security risks brought by the admission of third-party apps. Although the mobile OS security has received extensive research attention \cite{enck2011study,shabtai2010google,faruki2014android} and the security of prevalent mobile OSes has been continually hardened, the attack-defense race will continue and the emerging edge OSes may not be incorporated with the latest security enhancements that have been adopted by the mobile OSes.

\begin{figure}
  \centering
  \includegraphics[width=0.48\textwidth]{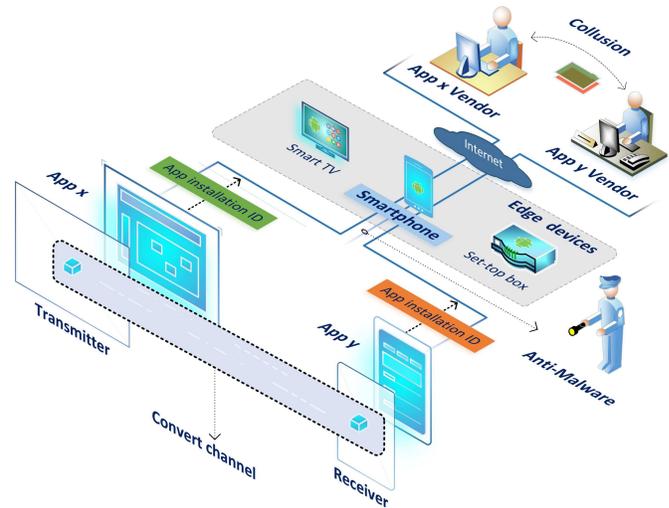}
  \caption{Covert device association among colluding apps on mobile or edge OSes (e.g., Android and Android Things).}
  \label{fig:association}
\end{figure}

The mobile and edge OSes mostly allow the user to run multiple apps concurrently. The communications among different apps should be performed over the overt channels. For instance, in Android, the inter-app interactions should be implemented using Android APIs such that the Android OS can monitor such communications \cite{enck2014taintdroid} and keep the user's awareness if needed. The covert communications among apps that the OS is unaware of will present a serious threat on the user's security and privacy \cite{memon2015colluding}. In this paper, we study a minimal form of inter-app covert communication called {\em covert device association}, which is illustrated by Fig.~\ref{fig:association}. Specifically, the vendors of two apps collude stealthily. Any two installations of their apps, referred to as $x$ and $y$, can determine whether they run on the same device. If yes, they are associated and can exchange massive data over the wideband {\em Internet channel} consisting of $x$'s and $y$'s connections to their cloud backends as well as the communication channel between the two colluding vendors.

The covert device association capability is a stepping stone for a number of further attacks that can be launched by the associated $x$ and $y$, including but not limited to the following:

\vspace{.2em}
\noindent {\bf De-anonymization:} If the user provides a meaningful identity (e.g., e-mail address or phone number) to $x$, the $y$, to which the user wishes to keep anonymous, can obtain the user's identity from the vendor of $x$ over the Internet channel. Particularly, in this paper, we will show that the $y$ can be a web session running in a mobile Tor browser \cite{tor-browser} for anonymity and it can establish the covert association with a local app $x$.

\vspace{0.2em}
\noindent {\bf Privilege escalation:} Current mobile OSes feature meticulously designed permission control mechanisms to manage the information that can be accessed by individual apps. Edge OSes will likely adopt similar designs for edge security. However, through the Internet channel, the associated $x$ and $y$ can access the information that is defined by the union of $x$'s and $y$'s permission sets. Therefore, the privileges of both $x$ and $y$ are stealthily escalated beyond the awareness of the victim edge OS.

The above follow-up attacks launched based on the covert device association can occur at large scales when the two apps are installed on many edge devices. Therefore, the covert device association can be a widespread threat should the edge OSes and the corresponding app distribution systems prevail.

In this paper, we investigate the covert device association via the edge processor workload. Existing studies \cite{claudio2012ACSAC,chandra2014scc,bhandari2017android,Jean2013ICA} have shown the feasibility of various covert channels via touchscreen brightness, vibration actuation and sensing, and etc., on mobile OSes. However, they focus on the local data exchange collusion between two apps that may require non-trivial synchronization mechanisms. Moreover, some covert channels (e.g., touchscreen) are not ubiquitous on a wider spectrum of edge platforms. Therefore, although the local data exchange collusion via the identified covert channels has been shown feasible, they are inefficient and may not be always available. Differently, in this paper, we show via extensive experiments that the processor workload provides a reliable covert channel that is ubiquitous on edge hardware platforms. Moreover, the studied covert device association requires a simple synchronization mechanism (i.e., scheduled rendezvous) and minimizes the usage of the covert channel by moving the data exchange collusion to the cloud backends. In summary, the high availability, the simple synchronization, and the minimal usage of the processor workload covert channel make the covert device association a credible and immediate security threat against the emerging edge computing systems.

In this paper, we present the prototype implementations of the covert device association via processor workload in three configurations: (i) two local apps on Android smartphone; (ii) a local app and a web session in a Tor browser \cite{tor-browser} on Android smartphone; and (iii) two local apps on a Raspberry Pi single-board computer running Android Things OS 1.0 \cite{android-things}. The implementations exploit several synthetic files that store information relevant to the processor's utilized cycles and the dynamic voltage and frequency scaling (DVFS). The accesses to these files do not require privileged permissions. We also present the results of evaluating the implementations including the impact of interfering computing tasks on the covert channel and smartphone's heat generation caused by the modulation of bits using processor workload. Results show that, the processor workload-based covert channel gives at least $0.25\,\text{bps}$ data rate with zero empirical bit error rate and maintains the smartphone temperature below human body temperature. With the $0.25\,\text{bps}$ data rate, the covert device association can be completed within 3.2 minutes.

The remainder of this paper is organized as follows. Section~\ref{sec:related} reviews related work. Section~\ref{sec:overview} overviews the covert device association threat. Section~\ref{sec:impl} presents prototype implementations. Section~\ref{sec:expri} presents evaluation results. Section~\ref{sec:discuss} discusses countermeasures. Section~\ref{sec:conclusion} concludes this paper.


\section{Related Work}
\label{sec:related}

{\em Device fingerprinting} and {\em covert channel} are two topics related to the device association threat studied in this paper. This section reviews the related work in these two topics.

Device fingerprinting uses device-unique metrics to identify and track devices. In \cite{bojinov2014mobile}, the accelerometer measurements of a smartphone are used to construct a hardware fingerprint. The study shows that the entropy of the fingerprint is sufficient to uniquely identify a device among thousands of devices with a low probability of collision.

In \cite{das2014you}, smartphones' microphones and speakers are shown to provide hardware fingerprints.

The study \cite{ba2018abc} shows that a picture captured by a smartphone camera contains the camera's fingerprint. Most of such fingerprinting techniques are developed for sensor-rich smartphones. The wider spectrum of edge devices may not be equipped with the used sensors. Moreover, when such fingerprinting techniques are used for device association, the two colluding apps $x$ and $y$ need to access the same fingerprint bearing sensors. The fingerprint extraction and match processes often incur considerable compute overhead. In contrast, the covert device association technique in this paper uses the processors that are pervasively available on any edge devices. The process of demodulating the processor workload to information bits involves light computation only.

A covert channel is a communication channel that is not by design and not managed by the OS. The covert channel between two processes running on the same device is often based on a shared resource. Memory access-based covert channels between two virtual machines in the cloud computing environment have been studied \cite{xiao2012covert,pessl2016drama,wu2014whispers}. On mobile devices, hardware accesses (e.g., Bluetooth, vibrator and accelerometer, screen brightness) have been exploited to establish covert channels \cite{claudio2012ACSAC,chandra2014scc,bhandari2017android,Jean2013ICA}. However, the used hardware components (e.g., vibrator, accelerometer, screen, etc) are not ubiquitously available on a wider spectrum of edge devices. Moreover, the accesses to the hardware components generally require certain permission.

Several studies exploit processors to establish covert channels \cite{MastiThermal,bartolini2016capacity,claudio2012ACSAC,chandra2014scc}. In \cite{MastiThermal,bartolini2016capacity}, the heat transfers between two cores on a server-class processor are used for the covert communications between two virtual machines running on the two cores. As virtualization is not widely adopted on edge devices, our approach uses the processor-wide workload as the medium of covert communications and does not resort to heat. This improves the efficiency of the covert channel and helps avoid high device temperatures that undermine the stealthiness of the attack on hand-held devices. The studies \cite{claudio2012ACSAC,chandra2014scc} mention the use of mobile processor utilization among various media for covert communications. However, these studies \cite{MastiThermal,bartolini2016capacity,claudio2012ACSAC,chandra2014scc} focus on covert communications only that require non-trivial synchronization mechanisms and fall short of addressing realistic factors such as interfering computing tasks. Differently, we focus on covert device association, a minimal form of inter-app covert communication that presents immediate threats to the edge computing systems, and perform evaluation under realistic settings.

The covert channel between two apps running on different devices is based on some physical signal that can propagate in space, e.g., sound \cite{deshotels2014inaudible}, heat \cite{Guri2015BitWhisper}, and electromagnetic radiation \cite{guri2015gsmem}. As smartphones' audio sub-system (loudspeaker and microphone) can work in a near-ultrasonic frequency band that cannot be perceived by human ears, ultrasound-based cross-device covert channel has been designed \cite{deshotels2014inaudible}. Such channel can be exploited to track users and infer users' privacy. The study \cite{arp2017privacy} found 234 Android apps that are constantly detecting ultrasonic beacons emitted from devices installed by mall stores to reveal the smartphone owners' shopping behaviors.

The ultrasound-based covert channel among personal compute devices (e.g., smartphones, wearables, and tablets) and home appliances (e.g., smart TVs and cars) can also be used for user desire/preference analysis and targeted advertisement \cite{waddell2016your}.


\section{Overview of Covert Device Association}
\label{sec:overview}

This section presents an overview of the covert device association between two colluding apps. Section~\ref{subsec:threat-model} presents the threat model. Section~\ref{subsec:attack-scenarios} discusses three attack scenarios. Section~\ref{subsec:modulation} presents the modulation and demodulation schemes using edge processor workload.

\subsection{Threat Model}
\label{subsec:threat-model}

We consider two app vendors that collude to perform covert device association. Let $X$ and $Y$ denote the apps published by the two vendors, respectively; let $\mathcal{X}$ and $\mathcal{Y}$ denote the finite sets of $X$'s and $Y$'s installations on the Internet-connected edge devices, respectively. The covert device association determines whether any two installations $x \in \mathcal{X}$ and $y \in \mathcal{Y}$ are on the same edge device. While this paper focuses on presenting the threat with two colluding app vendors, the attack tactic can be easily extended to address the case where more than two app vendors collude, which will be discussed at the end of this section.

If the apps can obtain the edge devices' unique identifiers (IDs), the covert device association is straightforward. However, as devices' unique IDs have strong security and privacy implications, they should be only accessible via overt mechanisms managed by the OSes. The latest mobile OSes have implemented various restriction and protection mechanisms for mobile edge devices' unique IDs. For instance, Android~10 only allows the {\em device/profile owner} apps that have certain privileged permissions (i.e., the {\em special carrier permission} or the {\em read privileged phone state}) to access the non-resettable IDs such as International Mobile Equipment Identity (IMEI) and serial number. Android 10 returns randomized media access control (MAC) addresses to all apps that are not device owner apps. Thus, obtaining the edge devices' unique IDs does not form a covert mechanism for device association and will become unreliable through the course of the edge OSes' security enhancements.

In this paper, the basic idea is that the app installation $x$, performing as the {\em transmitter}, influences the edge device's processor workload to modulate $x$'s installation ID that is denoted by $I(x)$ and unique within $\mathcal{X}$, while the app installation $y$, performing as the {\em receiver}, reads the processor workload to detect and demodulate the app installation ID. Fig.~\ref{fig:association} illustrates the process. If the receiver $y$ can detect an app installation ID, it sends its own app installation ID $I(y)$ and the detected ID to the vendor $Y$. As $X$ colludes with $Y$ and knows all the app installation IDs of $\mathcal{X}$, the colluding vendors can figure out the $X$'s installation instance within $\mathcal{X}$ that has the detected installation ID. Thus, the covert device association is achieved.

We discuss several issues related to the implementation of the above threat.
\paragraph{App installation ID generation}
As the colluding vendors identify the app installations based on both the app name (i.e., $X$ or $Y$) and the installation ID, each vendor can manage its own installation ID database independently. That is, the IDs of the installations in $\mathcal{X}$ or $\mathcal{Y}$ are unique, while $x \in \mathcal{X}$ and $y \in \mathcal{Y}$ may have the identical installation ID. We follow the rationale of IPv6 to use six bytes (i.e., 48 bits) to represent installation ID. The adversary vendor can also use fewer bits based on the estimated quantity of the installations of its app to reduce the covert communication overhead. For instance, from a statistics of Google Play as of September 2019 \cite{google-play-statistics}, 4-byte app installation ID can be adopted for 99.986\% apps that have less than one billion installations, among a total of 281,427 apps. When an app installation is initialized for the first-time use on an edge device, it communicates with the respective vendor's cloud backend to request an unique installation ID. The cloud backend can generate unique IDs sequentially or randomly. The app installation stores the received ID to the local storage for future use.
\paragraph{Synchronization mechanism}
The threat can adopt a {\em scheduled rendezvous} scheme for the synchronization between the transmitter and the receiver. Specifically, the colluding vendors stipulate periodic time instants for the covert communications, e.g., 04:00:00.000am of every Monday. When the apps can access the local time, the rendezvouses can be scheduled at the time instants when the edge devices are likely idle to reduce the impact of the interfering compute tasks on the processor workload-based covert communications, e.g., during the sleep time of the mobile devices' users.
\paragraph{More than two vendors collude}
In a coalition of more than two colluding vendors, each vendor's app can take turn to be the transmitter at the stipulated time instants, whereas all other vendors' apps act as the receivers. As a result, the coalition can figure out the subset of the apps running on each individual edge device.

\subsection{Attack Scenarios}
\label{subsec:attack-scenarios}

This section discusses three possible scenarios of the covert device association attack. Their prototype implementations in the Android OS and Android Things OS \cite{android-things} will be presented and evaluated in Section~\ref{sec:impl} and Section~\ref{sec:expri}, respectively.

\subsubsection{Covert device association between two mobile apps}

In this scenario, the two apps $x$ and $y$ are native apps running on top of a mobile OS. The transmitter app $x$ modulates its installation ID by affecting the processor workload. The receiver app $y$ detects and demodulates the installation ID by monitoring the processor workload. In Section~\ref{sec:impl}, we will implement this scenario based on the Android OS. We will elaborate the detailed constraints imposed by the Android OS for the receiver $y$ to access the processor workload and our approach to overcome the constraints.

\subsubsection{Covert device association between a mobile app and a web session}

In this scenario, a JavaScript-instrumented web page running in a web browser on top of a mobile OS is the transmitter $x$ and a native app running in the mobile OS is the receiver $y$. The web session ID of $x$ can be generated by the remote web server dynamically when the web browser requests the web page. Section~\ref{sec:impl} will present the technical details of using JavaScript to affect the processor workload and modulate the web session ID. The web browser can be a Tor browser \cite{tor-browser} that integrates the Tor for anonymous communications over an overlay network and various modifications for enhanced anonymity. The user wishes to achieve guaranteed anonymity by using the Tor browser. Unfortunately, if the user is not anonymous to $y$, through the covert device association, the anonymity protection provided by the Tor browser will be breached since the colluding vendors that provide the JavaScript-instrumented web page and the app $y$, respectively, can associate the user's identity (e.g., e-mail address) provided to $y$ with the anonymous web session $x$ or the user's pseudonym provided to $x$.

\subsubsection{Covert device association between two edge apps}

We extend the first scenario to address the case with two native apps running in an edge OS. We envisage that indoor appliances will increasingly adopt edge OSes that can run multiple apps concurrently. Many smart appliances have adopted customized Android OSes, such as TVs, fridges, and washers. The recently released Android Things \cite{android-things} is a lightweight edge OS that caters better into low-power and resource-constrained edge devices. The prototype implementation of the covert device association on Android Things will suggest the credibility of the attack on a wide spectrum of edge devices.


\subsection{Modulation/Demodulation using Processor Workload}
\label{subsec:modulation}

This section presents the processor workload-based modulation and demodulation methods. As we will discuss in Section~\ref{sec:impl}, the {\em processor time load} and {\em processor frequency load} are two metrics characterizing the processor workload. The modulation and demodulation methods presented in this section are agnostic to the detailed processor workload metrics.

\subsubsection{Modulation}

Unipolar non-return-to-zero (NRZ) line code is used to encode the app installation ID. For modulation, we consider the amplitude-shift keying (ASK) and the frequency-shift keying (FSK), which are two basic band-pass modulation schemes widely used in digital communication systems. We will compare the performance of the ASK-based and FSK-based covert communications via extensive evaluation, in the presence of interfering compute tasks. We do not use the phase-shift keying (PSK) because it is hard to control the phase of the carrier (i.e., the processor workload) in the presence of interfering workload.

\begin{figure}
  \centering
  \subfigure[ASK]
  {
    \includegraphics{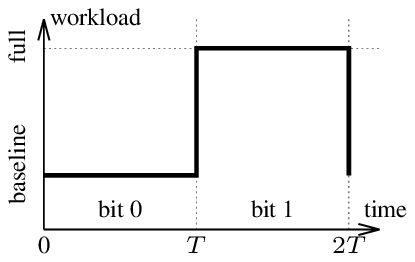}
  }
  \subfigure[FSK]
  {
    \includegraphics{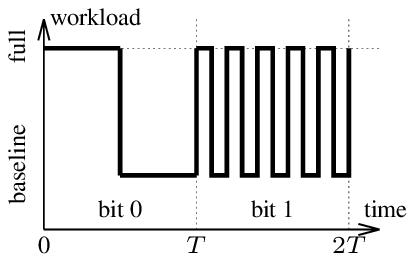}
  }
  \caption{Ideal processor workload waveforms of ASK and FSK.}
  \label{fig:waveforms}
\end{figure}

Fig.~\ref{fig:waveforms} illustrates the ideal processor workload waveforms of ASK and FSK.
Under ASK, a high processor workload represents binary 1, whereas a low processor workload represents binary 0. The two binaries use the identical {\em bit duration} that is denoted by $T$ in seconds. Our evaluation will vary the bit duration to achieve various data rates in bit per second (bps) and assess the resulted bit error rates (BERs).

Under FSK, the two binaries also use the identical bit duration $T$. The processor workload following a square wave with a period of $T$ seconds represents binary 0, whereas that following a square wave with a period of $\frac{T}{5}$ seconds represents binary 1. Thus, the carrier frequency for binary 1 is five times of the carrier frequency for binary 0. The frequencies are chosen to achieve well separation of the two states.

\subsubsection{Demodulation}

The receiver samples the processor workload to demodulate the signal. From the Nyquist sampling theorem, the sampling rate should be at least twice of the highest frequency of the carrier. We adopt an overprovisioned sampling rate of four times of the highest frequency of the carrier. Therefore, the sampling rates for the ASK and FSK schemes are $\frac{4}{T}\,\text{Hz}$ and $\frac{20}{T}\,\text{Hz}$, respectively.

Note that as the transmitter and the receiver are synchronized by the scheduled rendezvous mechanism, the demodulation can be coherent.

We use threshold-based bit detection. Specifically, for ASK, if 75\% of the sampled data points are higher than a detection threshold, the bit is detected as 1; otherwise, it is detected as 0. We use the average of the full processor workload and the baseline processor workload before the transmission of the installation ID as the detection threshold.

For FSK, we first perform the fast Fourier transform (FFT) on the sampled data over the current bit duration to generate the power spectrum density (PSD). If the peak value of the PSD occurs at a frequency higher than $\frac{3}{T}\,\text{Hz}$, the bit is detected as 1; otherwise, it is detected as 0. Note that the used frequency threshold of $\frac{3}{T}\,\text{Hz}$ is the mid point between the two carrier frequencies of $\frac{1}{T}\,\text{Hz}$ and $\frac{5}{T}\,\text{Hz}$ respectively representing binary 0 and 1.


\section{Prototype Implementations}
\label{sec:impl}

This section presents the prototype implementations of the three attack scenarios discussed in Section~\ref{subsec:attack-scenarios}. First, we present the device setup in Section~\ref{subsec:exp-setup}. Then, we present the processor workload metrics in Section~\ref{subsec:metrics}. Lastly, we present the implementation details of processor workload control and sensing in Section~\ref{subsec:control-sensing}.

\subsection{Device Setup}
\label{subsec:exp-setup}

Our implementations use three representative edge devices:
\begin{enumerate}
\item The edge device named {\bf Phone-A} is a Xiaomi Redmi 2 smartphone equipped with a 1.2GHz ARM Cortex-A53 quad-core processor that can run at eight different clock rates from 200MHz to 1.2GHz. Note that the run-time clock rate can be used to infer the processor workload, as discussed in Section~\ref{subsec:metrics}. The smartphone runs Android OS 5.1.1. It represents a low-cost smartphone according to the market retail price.
\item The edge device named {\bf Phone-B} is a Moto Z XT1650-03 smartphone that features ARM's big.LITTLE architecture consisting of two processors (a 1.8GHz dual-core Kryo processor and a 1.6GHz dual-core Kryo processor). It automatically switches between the two processors depending on run-time compute load. Such switching presents a challenge to the workload modulation and demodulation. The 1.8GHz processor can run at 21 different clock rates from 307MHz to 1.8GHz, while the 1.6GHz processor can run at 16 different clock rates from 307MHz to 1.6GHz. The phone runs Android 8.0.0 and represents a relatively high-end smartphone.
\item The edge device named {\bf RPi} is a Raspberry Pi 3 Model B single-board computer equipped with a 1.2GHz quad-core ARM8 processor. The processor can run at two different clock rates only, i.e., 600MHz and 1.2GHz. It runs Android Things 1.0.0.
\end{enumerate}

Our prototype implementations focus on Android OS, which has been widely adopted on smartphones and home appliances. The Android Things OS is a lightweight Android optimized for embedded devices. To be suitable for the devices without any user interface, the Android Things apps and their runtime permissions are deployed through over-the-air updates via a console. Multiple apps can be launched by the Android Things OS via a broadcast intent (BOOT\_COMPLETED) and run concurrently in the background.

\subsection{Processor Workload Metrics}
\label{subsec:metrics}

In this paper, we adopt two metrics to characterize the processor workload: {\em time load} and {\em frequency load}. In the following, we present how to obtain these metrics and the mobile OSes' restrictions on accessing these metrics.

\subsubsection{Time load}

The synthetic file ``/proc/stat'' of the Android OS and Android Things provides the accumulated and aggregated core and processor times in the unit of ten milliseconds spent for different types of processes (i.e., user, system, and idle times, etc). Thus, by reading this file twice at the start and end of a time duration, we can compute the percentage of the processor time spent for the user apps. This percentage is referred to the {\em time load} of the processor. Note that the Android or Android Things apps can also invoke the ``top'' utility to obtain the processor times. In fact, the ``top'' utility also relies on the information contained in ``/proc/stat''. Reading ``/proc/stat'' suffices for sensing time load.

\subsubsection{Frequency load}
Most modern processors feature DVFS that allows the processor clock rate to vary according to the real-time compute load and/or the configurations specified via Android's CPU governor. DVFS is critical to extend the battery lifetime of battery-based edge devices such as smartphones. Due to DVFS, the processor clock rate is an indicator of the processor workload. Although DVFS may have time delays in reacting to the variations of processor workload, the delay will not affect the information bit demodulation if the bit duration is long enough.
 
The maximum, minimum, and real-time processor clock rates can be obtained by reading the files with a prefix ``scaling\_'' in the ``/sys/devices/system/cpu/cpu$i$/cpufreq'' directory, where the $i$ represents the index of the core. The user app can read these files without any privileged permissions. For a device with $n$ cores, the frequency load is computed as:
\begin{equation*}
  \text{frequency load} = \frac{1}{n} \sum_{i=1}^{n} \frac{\text{current clock rate of the $i$th core}}{\text{max clock rate of the $i$th core}}.
\end{equation*}
Note that in this paper, we do not consider hyperthreading that virtualizes a physical core into multiple virtual cores, since this technique is not widely adopted for edge processors.

   \subsection{Processor Workload Control and Sensing}
   \label{subsec:control-sensing}

   This section presents the details of implementing the processor workload control and sensing for the modulation and demodulation processes presented in Section~\ref{subsec:modulation}, under the three attack scenarios.

   \subsubsection{Implementation for two Android apps}
   \label{subsubsec:impl-two-apps}

For processor workload control, to achieve a high workload level used by ASK or FSK, the transmitter app creates $n-2$ threads (where $n$ is the number of cores) and uses repeated matrix multiplications in each thread to increase the workload as much as possible until the high workload level should end. The rationale of using $n-2$ cycle-exhausting threads is as follows. As the Linux kernel tends to assign the threads evenly to the processor cores, our approach will most likely leave two cores not assigned with the cycle-exhausting threads. Thus, these two remaining cores can handle other compute tasks and retain the responsiveness of the OS to the user's input. Otherwise, the other compute tasks may severely interfere with the modulation/demodulation processes; the deteriorated system responsiveness to user's inputs can also reduce the stealthiness of the attack. Note that an Android app can obtain the value of $n$ by using the API \texttt{java.lang.Runtime.availableProcessors()}.

For processor workload sensing, with Android prior to version 8.0, the synthetic file ``/proc/stat'' can be read by user apps without requiring any privileged permissions. Thus, the time load-based demodulation can be implemented. However, with Android version 8.0 and newer, the SELinux configuration does not allow the user apps to access the synthetic file. Thus, only the frequency load-based demodulation can be implemented in Android 8.0 and newer.

\begin{figure}
  \centering
  \subfigure[Time load, ASK]{
    \includegraphics{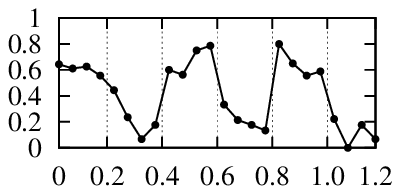}
  }
  \subfigure[Time load, FSK]{
    \includegraphics{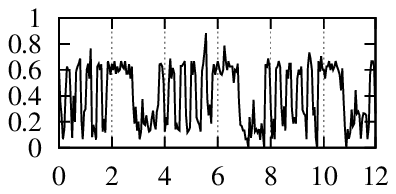}
  }
  \subfigure[Frequency load, ASK]{
    \includegraphics{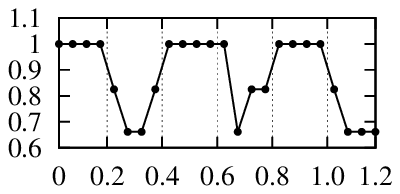}
  }
  \subfigure[Frequency load, FSK]{
    \includegraphics{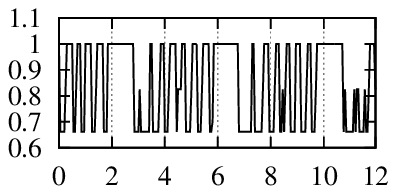}
  }
  \caption{The sensed workload traces of the first attack scenario (two Android apps) on Phone-A. The $x$-axis is time in seconds; the $y$-axis is the workload. The original data bits are `101010'. A movie is played during transmission to create interference.}
  \label{fig:s1-trace}
\end{figure}

Fig.~\ref{fig:s1-trace} shows the sensed workload traces under various combinations of modulation methods and processor workload metrics on Phone-A. During the transmissions, a movie is played to create interference. Although the waveforms are different from the ideal waveforms, all the shown waveforms can be correctly demodulated. From Fig.~\ref{fig:s1-trace}, we can also see that the frequency load traces are less noisy than the time load traces, due to the discrete nature of the frequency load.

\subsubsection{Implementation for Android app and web app}
\label{subsubsec:impl-native-web}

We use a JavaScript program that runs in a web browser in the Android OS to control the processor workload, while a native Android app as described in Section~\ref{subsubsec:impl-two-apps} senses and demodulates the processor workload. JavaScript was originally designed as a single-threaded scripting language. The latest JavaScript can use the Web Worker API of HTML5 to implement multi-threaded concurrency. By using Web Worker, we implement the modulation approach described in Section~\ref{subsubsec:impl-two-apps} by exhausting $n-2$ cores. Note that JavaScript can obtain the value of $n$ by accessing the read-only property \texttt{navigator.hardwareConcurrency}.

We also investigate whether the covert device association threat is valid against the Tor browser \cite{tor-browser}. Only the following two features of the Tor browser affect our design. First, to preserve the user's location privacy, the Tor browser provides Coordinated Universal Time (UTC) only. Thus, the scheduled rendezvous cannot be based on the local time and cannot exploit the night time to increase the stealthiness of the attack. Nevertheless, as the Tor browser unlikely runs in the background unattended for long periods of time (e.g., days) due to the mobile user's heightened caution in using anonymous communications, the opportunities for exploiting unattended times for attack stealthiness are limited. From our evaluation results in Section~\ref{sec:expri}, although the user activities may degrade the performance of the covert device association, the transmissions are still successful when the data rate is low. Thus, this restriction of the Tor browser does not invalidate the covert device association threat.

\begin{figure}
  \subfigure[Time load, ASK]{
    \includegraphics{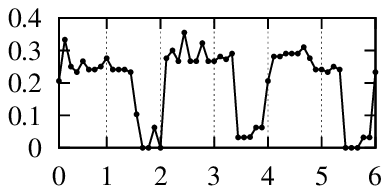}
  }
  \subfigure[Frequency load, ASK]{
    \includegraphics{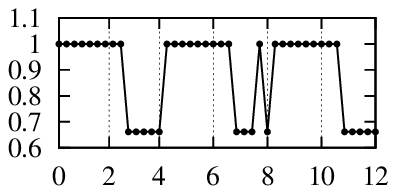}
  }
  \caption{The sensed workload traces of the second attack scenario (Android app and Tor browser) on Phone-A. The $x$-axis is time in seconds; the $y$-axis is the workload. Original data bits are `101010'. A movie is played to create interference.}
  \label{fig:s2-trace}
\end{figure}

Second, as recent studies \cite{schwarz2017fantastic,van2015clock,bosman2016dedup} show that the attackers can leverage the 1-millisecond accurate timer of JavaScript to de-anonymize the Tor browser, the Tor browser has reduced the JavaScript timer's resolution to 100 or 250 milliseconds \cite{perry2015bug}. As the processor workload modulation requires the JavaScript timer, the reduced timer resolution may lead to data rate reduction of the covert communications. However, from our evaluation results in Section~\ref{sec:expri}, the main bottleneck in modulation is the response delay of processor workload to the JavaScript program, rather than the reduced timer resolution. Fig.~\ref{fig:s2-trace} shows the sensed workload traces when an Android app and a web page in Tor browser running on Phone-A perform covert communications. A vertical dash line represents the boundary between two bit periods. We can see that, under ASK, there is a time delay of up to 0.5 seconds from when the JavaScript intends to reduce the workload to when the workload starts dropping. This is the aforementioned main limiting factor for JavaScript to achieve high data rate.

\subsubsection{Implementation for two Android Things 1.0 apps}
\label{subsubsec:impl-things}

\begin{figure}
  \subfigure[Time load, ASK]{
    \includegraphics{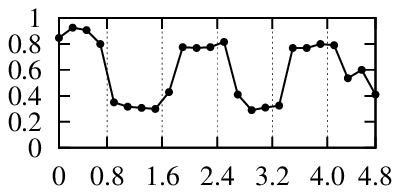}
  }
  \subfigure[Time load, FSK]{
    \includegraphics{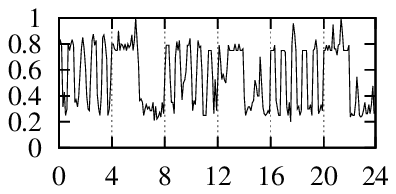}
  }
  \caption{The sensed workload traces of the third attack scenario (two Android Things 1.0 apps) on RPi. The $x$-axis is time in seconds; the $y$-axis is the workload. Original data bits are `101010'. A large file is compressed to create interference.}
  \label{fig:s3-trace}
\end{figure}

Our implementations for two Android 8.0 can be easily ported to Android Things 1.0. Oriented to embedded edge devices, the Android Things 1.0 does not prevent the user apps from accessing the synthetic files storing run-time processor time and frequency information as mentioned in Section~\ref{subsec:metrics}. Therefore, the receiver app can implement the demodulation methods based on either time workload or frequency workload. However, we find that the RPi's processor mostly operates at its maximum clock rate of 1.2GHz among its two available clock rates of 600MHz and 1.2GHz, even when it is idle. This is because the RPi design is not optimized for power conservation, rendering the frequency load-based modulation/demodulation ineffective. Differently, the battery-based smartphones actively scale down the clock rate to save battery energy when the processor is idle or less busy. Thus, in this paper, we only evaluate the time load-based scheme for this attack scenario. Fig.~\ref{fig:s3-trace} shows the sensed time load traces when the two Android Things apps communicate. During the transmissions, the RPi compresses a 300MB data file to introduce interference. Both ASK and FSK work satisfactorily.


\section{Benckmark Experiments}
\label{sec:expri}

We conduct experiments to benchmark our implementations for the three attack scenarios. Section~\ref{subsec:exp-settings} presents the experiment settings. Section~\ref{subsec:benchmark-results} presents the benchmark results.

\subsection{Experiment Settings and Methodology}
\label{subsec:exp-settings}

We deploy our implementations to the three hardware setups described in Section~\ref{subsec:exp-setup}. For each covert transmission, the transmitter generates a random sequence of 100 bits and then applies ASK or FSK to modulate the processor workloads. The receiver demodulates the sensed processor workload into bits and computes BER based on ground truth. We vary the bit duration $T$ to achieve different data rate and use the curve of BER versus data rate as the main covert communication performance profile. For each data rate setting, we perform four transmissions and use an error bar to present the minimum, maximum, and average BERs. We evaluate our implementations in the absence and presence of interfering computation. On the smartphones, we introduce the interfering computation by movie playback. During the movie playback, we also continuously create screen touch events to introduce uncertainties. On the RPi loaded with Android Things 1.0, we run a BASH script through the Android Debug Bridge (adb) to repeatedly compress a 300MB file using gzip to introduce interference. Note that this interference involves intensive computing and massive I/O operations with the microSD card.

\subsection{Benchmark Results}
\label{subsec:benchmark-results}

\begin{figure}
  \centering
  \subfigure[Time load, ASK]
  {
    \includegraphics{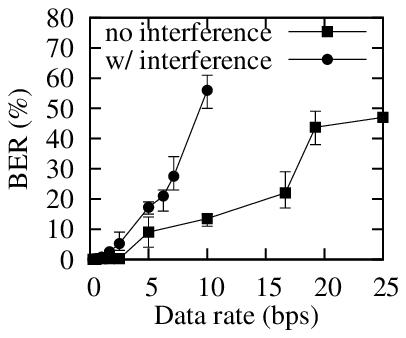}
  }
  \subfigure[Time load, FSK]
  {
    \includegraphics{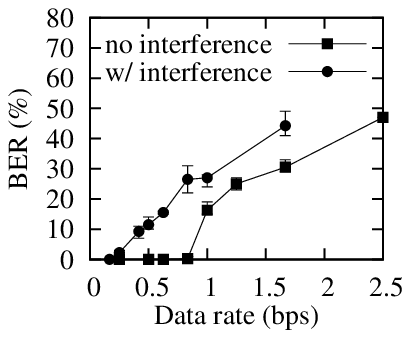}
  }
  \subfigure[Frequency load, ASK]
  {
    \includegraphics{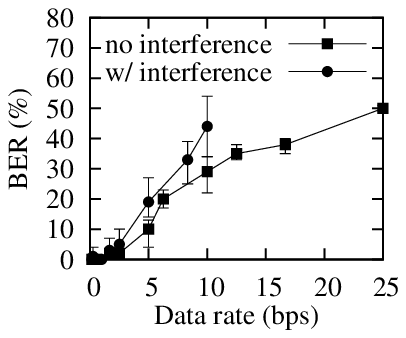}
  }
  \subfigure[Frequency load, FSK]
  {
    \includegraphics{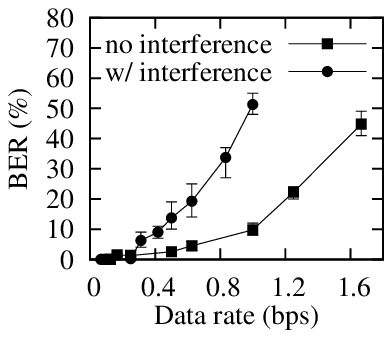}
  }
  \caption{BER vs. data rate on Phone-A. The subfigure captions specify the used workload metric and modulation method.}
  \label{fig:ber-vs-br-xiaomi}
\end{figure}

We present the three attack scenarios' benchmark results.

\subsubsection{Between two Android apps}

We conduct experiments using Phone-A. Fig.~\ref{fig:ber-vs-br-xiaomi} shows the BER versus data rate under various settings. We have the following three general observations. First, the BER increases with the data rate. Second, in the presence of interfering computation, the BER-versus-data rate curves become higher, because the interference causes bit errors. Third, compared with ASK, FSK achieves much lower data rates, because FSK uses multiple workload cycles to represent a bit. In addition, we have the following observation when we compare the subfigures of Fig.~\ref{fig:ber-vs-br-xiaomi}. The gaps between the two curves in Fig.~\ref{fig:ber-vs-br-xiaomi}(a) and Fig.~\ref{fig:ber-vs-br-xiaomi}(d) are larger than those in Fig.~\ref{fig:ber-vs-br-xiaomi}(b) and Fig.~\ref{fig:ber-vs-br-xiaomi}(c). This suggests that the demodulation methods based on the combinations of $\langle\text{time load}, \text{FSK}\rangle$ and $\langle\text{frequency load}, \text{ASK}\rangle$ are more resilient to interference than those based on $\langle\text{time load}, \text{ASK}\rangle$ and $\langle\text{frequency load}, \text{FSK}\rangle$.

Then, we investigate the heat generation during the covert transmission process, since overheating may cause the mobile user's suspicion and investigation. In our experiments, we record the smartphone's battery temperature that can characterize the overall heat generation status of the smartphone. During the experiments, the phone's battery temperatures are around 35\textdegree{}C only, lower than the human body temperature. Therefore, the covert transmission does not lead to salient overheating.

\begin{figure}
  \centering
  \subfigure[Frequency load, ASK]
  {
    \includegraphics{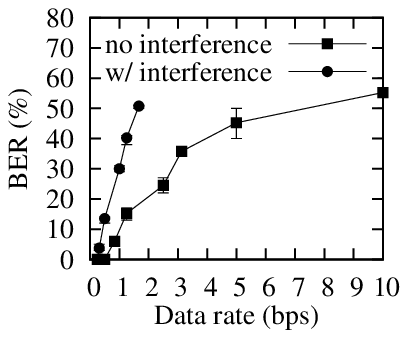}
  }
  \subfigure[Frequency load, FSK]
  {
    \includegraphics{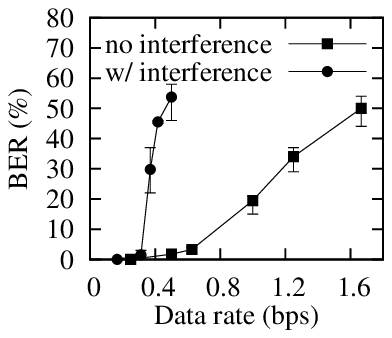}
  }
  \caption{BER vs. data rate on Phone-B.}
  \label{fig:ber-vs-br-moto}
\end{figure}

Lastly, we conduct experiments using Phone-B. A key difference between Phone-A and Phone-B is that Phone-B is equipped with two processors (i.e., ARM's big.LITTLE architecture). The phone can dynamically switch between the two processors depending on the real-time compute workload. This introduces challenges to maintaining the processor workload for modulation purpose. In addition, as Phone-B runs Android 8.0, it cannot measure the time load due to the OS' restriction as discussed in Section~\ref{subsubsec:impl-two-apps}. Fig.~\ref{fig:ber-vs-br-moto} shows the experiment results on Phone-B, where the frequency load is used as the processor workload metric. By comparing the results with those shown in Figs.~\ref{fig:ber-vs-br-xiaomi}(c) and \ref{fig:ber-vs-br-xiaomi}(d) for Phone-A, the covert transmissions on Phone-B in general have higher BERs for the same data rates. This degradation of covert communication performance is due to the extra uncertainty introduced by the dynamic switching between the two processors. Nevertheless, as long as the data rate is no greater than $0.25\,\text{bps}$, the empirical BERs measured under the various settings are zero.

\subsubsection{Between Android app and web app}

\begin{table}
  \centering
  \caption{Maximum achievable data rates in bps without any bit errors in the absence and presence of interference.}
  \label{tab:max-data-rate}
  \begin{tabular}{c|c|c|c|c}
    \Xhline{1.2pt}
    Browser & \multicolumn{2}{c|}{Google Chrome} & \multicolumn{2}{c}{Tor browser} \\
    \hline
    & \multicolumn{4}{c}{} \\
    \multirow{2}{*}{Setting} & \multicolumn{4}{c}{ASK, no interference} \\
    \cline{2-5}
    & time load & frequency load & time load & frequency load \\
    \hline
    Max bps & 0.63 & 0.89 & 0.64 & 0.91 \\
    \Xhline{1.2pt} \\
    \multirow{2}{*}{Setting} & \multicolumn{4}{c}{ASK, with interference} \\
    \cline{2-5}
    & time load & frequency load & time load & frequency load \\
    \hline
    Max bps & 0.61 & 0.81 & 0.48 & 0.72 \\
    \Xhline{1.2pt}
  \end{tabular}
\end{table}

We evaluate the implementation presented in Section~\ref{subsubsec:impl-native-web} using both the Google Chrome and the Tor browser on Phone-A. As shown in Section~\ref{subsubsec:impl-native-web}, the processor workload has delays in responding to the JavaScript program's intent to change the processor's workload. This can be caused by the browser's design for controlling the JavaScript engine's use of compute resources. As such, when the bit duration is too short, the modulation cannot be performed correctly. Table~\ref{tab:max-data-rate} shows the maximum achievable data rates with zero empirical BER in the absence and presence of interference. When the data rates are higher than the values shown in Table~\ref{tab:max-data-rate}, the modulation cannot be performed correctly. Moreover, the results shown in Table~\ref{tab:max-data-rate} for the Google Chrome and the Tor browser are similar. The results obtained in the absence and presence of interference are also similar. These results suggest that the bottleneck preventing from achieving higher data rates is the processor workload's delay in responding to the JavaScript program, rather than the Tor browser's reduced timer resolution and the impact of interference.

\subsubsection{Between two Android Things 1.0 apps}

\begin{figure}
  \centering
  \subfigure[Time load, ASK]
  {
    \includegraphics{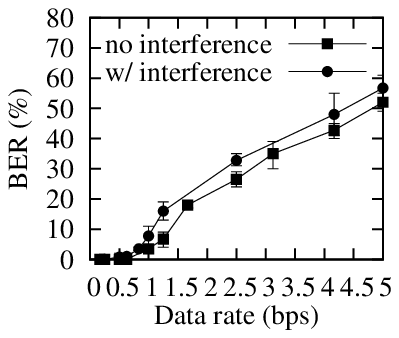}
  }
  \subfigure[Time load, FSK]
  {
    \includegraphics{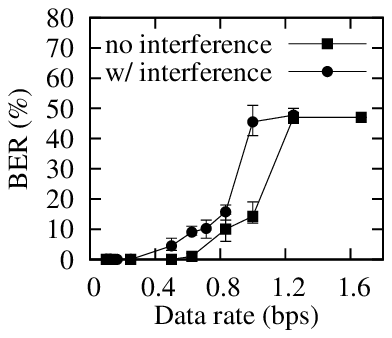}
  }
  \caption{BER vs. data rate on RPi running Android Things 1.0.}
  \label{fig:ber-vs-dr-pi}
\end{figure}

We evaluate our implementation for two Android Things 1.0 apps running on the RPi. As discussed in Section~\ref{subsubsec:impl-things}, RPi's limited frequency scales and inactive DVFS render the frequency load an ineffective processor workload metric. Thus, we only evaluate the time load-based covert transmissions. Fig.~\ref{fig:ber-vs-dr-pi} shows the results. We can see that, in both subfigures, the gap between the two curves corresponding to the presence and absence of interference is small. This suggests that the covert transmissions are resilient to the interference.

\subsubsection{Summary}

From the benchmark results obtained on the three edge devices, the covert transmission via processor workload can achieve a data rate of at least $0.25\,\text{bps}$ with zero empirical BER under the three attack scenarios, without requiring any privileged permissions. If the app installation ID is represented using six bytes, the covert device association can be completed within $6 \times 8 / 0.25\,\text{seconds} = 3.2\,\text{minutes}$.


\section{Discussion on Countermeasures}
\label{sec:discuss}

As discussed in Section~\ref{subsec:control-sensing}. Android 8.0 and newer have restricted the access to the synthetic file containing processor times. The latest Android (i.e., 10.0) still allows the access to the synthetic file containing DVFS information. Thus, the considered threat is valid on all current Android devices. Security enhancements can impose permission requirement for accessing the file containing DVFS information. However, more advanced workload sensing technique can be developed. For example, the receiver can measure the computation time of a certain compute-intensive task to infer the processor workload. As such, the receiver does not need to access any synthetic file. A possible further defense can be reducing the resolution of the timer APIs provided to apps. However, this defense may affect the functionality of time-critical apps. In summary, it is challenging to develop an ideal countermeasure against the studied threat.

While the main purpose of this paper is to arouse the awareness of the potential widespread covert device association attack, we also would like to stress that effective countermeasures against the threat are still lacking now and important for future work. The design of the countermeasures should well balance various factors such as the defense effectiveness, the compute overhead, the battery energy consumption on mobile devices, the impact on the functionality of the apps, and etc.


\section{Conclusion}
\label{sec:conclusion}

In this paper, we investigated a minimal form of inter-application covert communication called covert device association in the context of edge computing systems. The considered threat can widely spread and can be used as a stepping stone to launch other attacks such as de-anonymization and privilege escalation. It exploits a reliable and ubiquitous covert channel based on the edge device's processor workload. We implemented three attack scenarios, i.e., the covert device associations between two Android apps, between an Android app and a mobile web app, and between two Android Things 1.0 apps. Results show that the processor workload channel gives at least 0.25 bps data rate and the covert device association can be completed within 3.2 minutes.


\IEEEtriggeratref{21}

\bibliographystyle{IEEEtran}
\bibliography{bibliography}

\end{document}